# Matrix order differintegration

Mark Naber [a]

Department of Mathematics
Monroe County Community College
Monroe, Michigan 48161-9746

**ABSTRACT**

The Riemann-Liouville formula for fractional derivatives and integrals (differintegration) is used to derive formulae for matrix order derivatives and integrals. That is, the parameter for integration and differentiation is allowed to assume matrix values. It is found that the computation of derivatives and integrals to matrix order is well defined for any square matrix over the complex numbers. Some properties are worked out for special classes of matrices. It is hoped that this new formalism will be of use in the study of systems of fractional differential equations and sequential fractional differential equations.

[a] mnaber@monroeccc.edu

# 1. Introduction

The concepts of integration and differentiation have been expanded many times throughout the development of calculus. Almost immediately after the formulation of classical calculus by Leibnitz and Newton the question of half order derivatives arose (see [7] and [8] for historical reviews). Derivatives of complex and purely imaginary order were later developed (see e.g. [6] and references there in). Phillips considered fractional order derivatives of matrix-valued functions with respect to their matrix arguments, fractional matrix calculus. These operations were found to be well defined and subsequently applied to econometric distribution theory (see [4] and [5]). The applications of fractional calculus are numerous and the interested reader should consult [7], [8], and [9] for details.

In this paper the formulae from the Riemann-Liouville formulation of fractional calculus are adapted to matrix order for any matrix $A \in C^{n \times n}$ ($n \times n$ square matrices over the complex numbers). It is hoped that this new formalism will be of use in the study of systems of fractional differintegral equations and sequential fractional differential equations. Fractional differintegral equations have found a wide variety application particularly in fractional order dynamical systems (see [9] for more examples and other references).

In the following section the basic formulae from the Riemann-Liouville fractional calculus are given as well as several identities that will be useful for the matrix order material. In sec. 3, properties of functions with matrix arguments are reviewed. This sets notation and provides a convenient reference for reader. Sec. 4 gives the definition for matrix order derivatives. Notation is given as well as some new formula for derivatives of fractional derivatives. In sec. 5 properties for the matrix order derivatives are worked out.

# 2. Some Formula from Fractional Calculus

In this section the basic formulae of Riemann-Liouville fractional calculus are given, fractional integration and differentiation (differintegration). The reader should consult [7], [8], and [9] for further details and applications. $\Gamma(\lambda)$ is the gamma function (generalized factorial) of the parameter '$\lambda$' (i.e. $\Gamma(n+1) = n!$ for all whole numbers, n).

$$_aD_x^{-\lambda}f(x) = \frac{1}{\Gamma(\lambda)}\int_a^x \frac{f(\xi)d\xi}{(x-\xi)^{1-\lambda}}, \quad \text{Re}(\lambda) > 0, \tag{1}$$

$$_aD_x^{\lambda}f(x) = \frac{\partial^n}{\partial x^n}\left[\frac{1}{\Gamma(n-\lambda)}\int_a^x \frac{f(\xi)d\xi}{(x-\xi)^{\lambda-n+1}}\right], \quad \frac{\text{Re}(\lambda) \geq 0}{n > \text{Re}(\lambda) \text{ (n is whole)}}. \tag{2}$$

The parameter $\lambda$ is the order of the integral or derivative and is allowed to be complex. Eq. (1) is a fractional integral and eq. (2) is a fractional derivative. Taken together, the operation is referred to as differintegration (see page 61 of [7]). Using eqs. (1) and (2) an analytic function can be defined (see page 49 of [7]) for a given f(x) that is suitably differentiable.

$$g(\lambda) = \begin{cases} _aD_x^{\lambda}f(x) = \dfrac{1}{\Gamma(-\lambda)}\int_a^x \dfrac{f(\xi)d\xi}{(x-\xi)^{1+\lambda}}, & \text{Re}(\lambda) < 0, \\[2ex] _aD_x^{\lambda}f(x) = \dfrac{\partial^n}{\partial x^n}\left[\dfrac{1}{\Gamma(n-\lambda)}\int_a^x \dfrac{f(\xi)d\xi}{(x-\xi)^{\lambda-n+1}}\right] & \dfrac{\text{Re}(\lambda) \geq 0}{n > \text{Re}(\lambda) \text{ (n is whole)}} \end{cases} \tag{3}$$

Equations (4) – (11) are some identities from fractional calculus, m is a whole number and p and q are complex numbers whose real part is greater than zero.

$$\frac{\partial^m}{\partial x^m} {}_aD_x^q f(x) = {}_aD_x^{q+m}f(x) \tag{4}$$

$$_aD_x^q {}_aD_x^{-q}f(x) = f(x) \tag{5}$$

$$_aD_x^{-q} {}_aD_x^q f(x) \neq f(x) \tag{6}$$

$$_aD_x^p {}_aD_x^{-q}f(x) = {}_aD_x^{p-q}f(x) \tag{7}$$

$$_aD_x^p {}_aD_x^q f(x) = {}_aD_x^{p+q}f(x) - \sum_{j=1}^{k} {}_aD_x^{q-j}f(x)\bigg|_{x=a} \frac{(x-a)^{-p-j}}{\Gamma(1-p-j)} \tag{8}$$

$$_aD_x^{-p}\,_aD_x^q f(x) = {_aD_x^{q-p}} f(x) - \sum_{j=1}^{k} {_aD_x^{q-j}} f(x)\Big|_{x=a} \frac{(x-a)^{p-j}}{\Gamma(1+p-j)} \qquad (9)$$

$$_aD_x^{-p}\left(_aD_x^{-q}(f(x))\right) = {_aD_x^{-(p+q)}}(f(x)) \qquad (10)$$

$$_aD_x^q(fg) = \sum_{j=0}^{\infty} \binom{q}{j} \left(_aD_x^{q-j} f(x)\right)\left(\frac{\partial^j g}{\partial x^j}\right) \qquad (11)$$

In eqs. (8) and (9) 'k' is the first whole number $\geq \mathrm{Re}(q)$. The above formula and definitions can be found in [7], [8], and [9].

## 3. Functions of Matrices

To develop the notion of matrix order derivatives and integrals (differintegrals) it is useful to review the properties of functions with matrix order arguments. Three different means of computing functions with matrix order arguments will be used in this paper. The method chosen will be determined by the properties of the matrix being used and the specific application being considered.

A given non-zero matrix $A \in C^{n \times n}$ can be placed into one or more of the following sets; Jordan block diagonalizable, diagonalizable, and/or normal. These sets will be denoted by $B_n$, $D_n$, and $N_n$, respectively. Note that $N_n \subset D_n \subset B_n$ and every non-zero matrix $A \in C^{n \times n}$ is at least Jordan block diagonalizable. Recall that the set of normal matrices contains the following subsets: real symmetric, real skew-symmetric, positive and negative definite, Hermitian, skew-Hermitian, orthogonal, and unitary (see page 548 of [2] for further details). If a matrix A is normal then there exists a unitary matrix U such that

$$A = UDU^*. \qquad (12)$$

Where D is a diagonal matrix with the eigenvalues of A as the entries and $*$ denotes conjugate transpose. If A is real and normal then equation (12) reduces to

$$A = ODO^T \qquad (13)$$

and O is an orthogonal matrix. If a matrix A is diagonalizable then there exists an invertible matrix P such that

$$A = PDP^{-1}. \quad (14)$$

Alternatively, if A is diagonalizable the spectral theorem (see page 517 of [2]) can be used to find another representation for A. If $A \in D_n$ then there exists a set of matrices $\{G_1, ..., G_k\}$ such that,

$$A = \sum_{i=1}^{k} G_i \lambda_i, \quad (15)$$

where $\{\lambda_1, ..., \lambda_k\}$ are the unrepeated eigenvalues of A. The matrices $\{G_1, ..., G_k\}$ have the following properties;

$$G_i G_j = 0 \text{ for } i \neq j, \quad (16)$$

$$G_i G_i = G_i, \quad (17)$$

$$I = \sum_{i=1}^{k} G_i. \quad (18)$$

If a matrix is Jordan block diagonalizable, then there is an invertible matrix P such that

$$A = PJP^{-1} = P \begin{pmatrix} J(\lambda_1) & & 0 \\ & \ddots & \\ 0 & & J(\lambda_k) \end{pmatrix} P^{-1} \quad (19)$$

Where $\lambda_1, ..., \lambda_k$ are the distinct eigenvalues of A ($k < n$), and $J(\lambda_j)$ are the Jordan segments for the eigenvalue $\lambda_j$.

$$J(\lambda_j) = \begin{pmatrix} \lambda_j & 1 & & \\ & \ddots & \ddots & \\ & & \ddots & 1 \\ & & & \lambda_j \end{pmatrix} \quad (20)$$

There are several ways to define a matrix function. Let $A \in N_n$ with eigenvalues $\lambda_1, ..., \lambda_n$ (they need not be distinct) and let g(z) be a function that is defined for all the eigenvalues of A. Then g(A) can be expressed as

$$g(A) = U \begin{pmatrix} g(\lambda_1) & & \\ & \ddots & \\ & & g(\lambda_n) \end{pmatrix} U^*. \tag{21}$$

If A is real replace U by O and $U^*$ by $O^T$. If A is diagonalizable then equation (21) would be written as,

$$g(A) = P \begin{pmatrix} g(\lambda_1) & & \\ & \ddots & \\ & & g(\lambda_n) \end{pmatrix} P^{-1}. \tag{22}$$

The spectral representation can also be used (in either case).

$$g(A) = \sum_{i=1}^{k} G_i \, g(\lambda_i) \tag{23}$$

Where $\lambda_1, \ldots, \lambda_k$ are the distinct eigenvalues. For matrices that are Jordan block diagonalizable but not in $D_n$ the situation is somewhat more complicated.

$$g(A) = P \begin{pmatrix} g(J(\lambda_1)) & & \\ & \ddots & \\ & & g(J(\lambda_k)) \end{pmatrix} P^{-1} \tag{24}$$

Where,

$$g(J(\lambda_i)) = \begin{pmatrix} g(\lambda_i) & g'(\lambda_i) & g''(\lambda_i)/2! & \cdots & g^{(m-1)}(\lambda_i)/(m-1)! \\ & g(\lambda_i) & g'(\lambda_i) & \cdots & g^{(m-2)}(\lambda_i)/(m-2)! \\ & & \ddots & \ddots & \vdots \\ & & & g(\lambda_i) & g'(\lambda_i) \\ & & & & g(\lambda_i) \end{pmatrix}. \tag{25}$$

$g(J(\lambda_i))$ is an m × m upper triangular matrix. Note that for this case the function need be at least differentiable of order m-1 at $\lambda_i$. For further details see [2].

## 4. Matrix Order Differintegration

To develop matrix order differintegration eq. (3) is utilized in conjunction with eqs. (21) - (25). Let $A \in N_n$ then the differintegral of order A is given by,

$$_aD_x^A = U\begin{pmatrix} _aD_x^{\lambda_1} & & \\ & \ddots & \\ & & _aD_x^{\lambda_n} \end{pmatrix} U^*. \tag{26}$$

If A is real and normal then,

$$_aD_x^A = O\begin{pmatrix} _aD_x^{\lambda_1} & & \\ & \ddots & \\ & & _aD_x^{\lambda_n} \end{pmatrix} O^T \tag{27}$$

If $A \in D_n$ then

$$_aD_x^A = P\begin{pmatrix} _aD_x^{\lambda_1} & & \\ & \ddots & \\ & & _aD_x^{\lambda_n} \end{pmatrix} P^{-1} \tag{28}$$

Any of the above three cases, eqs. (26), (27), or (28), can be expressed using the spectral theorem.

$$_aD_x^A = \sum_{i=1}^{k} G_i \; _aD_x^{\lambda_i} \tag{29}$$

Where the $G_i$ are the matrices as given by the spectral theorem and the sum is over the unrepeated eigenvalues.

For matrices that are Jordan block diagonalizable but not diagonalizable some notation must be introduced. Recall eq. (3) from sec. 1,

$$g(\lambda) = \begin{cases} _aD_x^\lambda f(x) = \dfrac{1}{\Gamma(\lambda)} \int_a^x \dfrac{f(\xi)d\xi}{(x-\xi)^{1-\lambda}}, & \text{Re}(\lambda) < 0, \\[1em] _aD_x^\lambda f(x) = \dfrac{\partial^n}{\partial x^n}\left[\dfrac{1}{\Gamma(n-\lambda)}\int_a^x \dfrac{f(\xi)d\xi}{(x-\xi)^{\lambda-n+1}}\right] & \dfrac{\text{Re}(\lambda) \geq 0}{n > \text{Re}(\lambda) \text{ (n is whole)}} \end{cases}. \tag{30}$$

Derivatives of $g(\lambda)$ with respect to $\lambda$ will appear in the expression for the Jordan matrices,

$$\dfrac{d^k}{d\lambda^k} g(\lambda) = \begin{cases} _a^k D_x^\lambda f(x) = \dfrac{d^k}{d\lambda^k}\dfrac{1}{\Gamma(\lambda)}\int_a^x \dfrac{f(\xi)d\xi}{(x-\xi)^{1-\lambda}}, & \text{Re}(\lambda) < 0, \\[1em] _a^k D_x^\lambda f(x) = \dfrac{d^k}{d\lambda^k}\dfrac{\partial^n}{\partial x^n}\left[\dfrac{1}{\Gamma(n-\lambda)}\int_a^x \dfrac{f(\xi)d\xi}{(x-\xi)^{\lambda-n+1}}\right] & \dfrac{\text{Re}(\lambda) \geq 0}{n > \text{Re}(\lambda) \text{ (n is whole)}} \end{cases}. \tag{31}$$

In eq. (31) k is restricted to be a whole number. The ${}_a^k D_x^\lambda$ can be expressed in terms of ${}_a D_x^\lambda$.

$$
{}_a^1 D_x^\lambda f(x) = -\left(\frac{d}{d\lambda}\ln(-\lambda) + \ln(x)\right) {}_a D_x^\lambda f(x) - \sum_{n=1}^{\infty} \frac{{}_a D_x^\lambda(x^n f(x))}{n\, x^n} \qquad \mathrm{Re}(\lambda) < 0 \qquad (32)
$$

$$
{}_a^1 D_x^\lambda f(x) = -\frac{\partial^n}{\partial x^n}\left(\left(\frac{d}{d\lambda}\ln(-\lambda) + \ln(x)\right) {}_a D_x^{\lambda-n} f(x) + \sum_{n=1}^{\infty} \frac{{}_a D_x^{\lambda-n}(x^n f(x))}{n\, x^n}\right) \quad \begin{array}{l}\mathrm{Re}(\lambda) \geq 0 \\ n > \mathrm{Re}(\lambda)\ (\text{n is whole})\end{array} \quad (33)
$$

Similar expressions are available for the higher order derivatives of ${}_a D_x^\lambda$. With this notation matrix order differintegrals for $A \in B_n$ are denoted by

$$
{}_a D_x^A = P \begin{pmatrix} {}_a D_x^{J(\lambda_1)} & & \\ & \ddots & \\ & & {}_a D_x^{J(\lambda_k)} \end{pmatrix} P^{-1} \qquad (34)
$$

where,

$$
{}_a D_x^{J(\lambda_i)} = \begin{pmatrix} {}_a^0 D_x^{\lambda_i} & {}_a^1 D_x^{\lambda_i}/1! & \cdots & {}_a^{m-1} D_x^{\lambda_i}/(m-1)! \\ & {}_a^0 D_x^{\lambda_i} & \cdots & {}_a^{m-2} D_x^{\lambda_i}/(m-2)! \\ & & \ddots & \vdots \\ & & & {}_a^0 D_x^{\lambda_i} \end{pmatrix} \qquad (35)
$$

## 5. Properties

To determine some of the properties of matrix order differintegrals consider the composition of two matrix order differintegrals. Let A and $B \in D^n$ such that $A = PDP^{-1}$ and $B = QEQ^{-1}$ where D and E are diagonal matrices with the eigenvalues of A and B as entries. Denote the eigenvalues of A as $\lambda_i$ and the eigenvalues of B as $\rho_i$. Then,

$$
{}_a D_x^A {}_a D_x^B = P \begin{pmatrix} {}_a D_x^{\lambda_1} & & \\ & \ddots & \\ & & {}_a D_x^{\lambda_n} \end{pmatrix} P^{-1} Q \begin{pmatrix} {}_a D_x^{\rho_1} & & \\ & \ddots & \\ & & {}_a D_x^{\rho_n} \end{pmatrix} Q^{-1}. \qquad (36)
$$

To simplify this denote by $R = P^{-1}Q$ and the components of R as $R_{ij}$. Eq. (36) now has the form,

$$_aD_x^A {}_aD_x^B = P\left[R_{ija} D_x^{\lambda_i} {}_aD_x^{\rho_j}\right] Q^{-1}. \tag{37}$$

There is no sum over the repeated indices, they merely denote the components of the matrix between P and $Q^{-1}$. Alternatively the spectral theorem can be used to obtain another representation for the right hand side of eq. (36).

$$_aD_x^A {}_aD_x^B = \sum_{i=1}^{k} G_i {}_aD_x^{\lambda_i} \sum_{j=1}^{l} H_j {}_aD_x^{\rho_j} = \sum_{i=1}^{k}\sum_{j=1}^{l} G_i H_j {}_aD_x^{\lambda_i} {}_aD_x^{\rho_j} \tag{38}$$

Where $A = \sum_{i=1}^{k} G_i \lambda_i$ and $B = \sum_{j=1}^{l} H_j \rho_j$.

Eq. (4) carries over to the matrix order case.

$$\frac{\partial^m}{\partial x^m} {}_aD_x^A f(x) = \frac{\partial^m}{\partial x^m} \sum_{i=1}^{k} G_i {}_aD_x^{\lambda_i} f(x) \tag{39}$$

$$= \sum_{i=1}^{k} G_i \frac{\partial^m}{\partial x^m} {}_aD_x^{\lambda_i} f(x) \tag{40}$$

$$= \sum_{i=1}^{k} G_i {}_aD_x^{\lambda_i + m} f(x) \tag{41}$$

$$= {}_aD_x^{A+mI} f(x) \tag{42}$$

I is the identity matrix, m is a whole number, and f(x) may be a scalar, vector, or matrix order function.

Eq. (5) carries over in the following sense, Let A be a matrix such that $Re(\lambda_i) \geq 0$. Now consider eq. (36) with B replaced by –A.

$$_aD_x^A {}_aD_x^{-A} = P \begin{pmatrix} {}_aD_x^{\lambda_1} & & \\ & \ddots & \\ & & {}_aD_x^{\lambda_n} \end{pmatrix} \begin{pmatrix} {}_aD_x^{-\lambda_1} & & \\ & \ddots & \\ & & {}_aD_x^{-\lambda_n} \end{pmatrix} P^{-1} \tag{43}$$

$$_aD_x^A {}_aD_x^{-A} = P \begin{pmatrix} {}_aD_x^{\lambda_1} {}_aD_x^{-\lambda_1} & & \\ & \ddots & \\ & & {}_aD_x^{\lambda_n} {}_aD_x^{-\lambda_n} \end{pmatrix} P^{-1} \tag{44}$$

Which, by eq. (5), reduces to the identity operator.

Suppose that A and B commute, then A and B can be diagonalized by the same matrix, say P. Eq. (36) is now,

$$_aD_x^A \,_aD_x^B = P \begin{pmatrix} _aD_x^{\lambda_1} \,_aD_x^{\rho_1} & & \\ & \ddots & \\ & & _aD_x^{\lambda_n} \,_aD_x^{\rho_n} \end{pmatrix} P^{-1}. \tag{45}$$

A further simplification of eq. (45) will occur using equations (5), (7), (8), (9), or (10) as the signs of the eigenvalues dictate. For example, if the eigenvalues of A and B are such that $\text{Re}(\lambda_i) \le 0$ and $\text{Re}(\rho_i) \le 0$ then eq. (45) becomes,

$$_aD_x^A \,_aD_x^B = P \begin{pmatrix} _aD_x^{\lambda_1+\rho_1} & & \\ & \ddots & \\ & & _aD_x^{\lambda_1+\rho_1} \end{pmatrix} P^{-1} = \,_aD_x^{A+B}. \tag{46}$$

If the eigenvalues of matrices A and B are such that $\text{Re}(\lambda_i) \le 0$ and $\text{Re}(\rho_i) \le 0$ but not commuting then eq. (36) is,

$$_aD_x^A \,_aD_x^B = P \left[ R_{ij} \,_aD_x^{\lambda_i+\rho_j} \right] Q^{-1}. \tag{47}$$

Or, using the spectral theorem,

$$_aD_x^A \,_aD_x^B = \sum_{i=1}^{k} \sum_{j=1}^{l} G_i H_j \,_aD_x^{\lambda_i+\rho_j}. \tag{48}$$

Let A, B $\in N_n$ with A = $PDP^T$ and B = $QEQ^T$ where D and E are diagonal. Denote the eigenvalues of A by $\lambda_i$ and the eigenvalues of B by $\rho_i$. Now consider the transpose of eq. (36).

$$\left(_aD_x^A \,_aD_x^B\right)^T = Q \begin{pmatrix} _aD_x^{\rho_1} & & 0 \\ & \ddots & \\ 0 & & _aD_x^{\rho_n} \end{pmatrix} Q^T P \begin{pmatrix} _aD_x^{\lambda_1} & & 0 \\ & \ddots & \\ 0 & & _aD_x^{\lambda_n} \end{pmatrix} P^T \tag{49}$$

$$= \,_aD_x^B \,_aD_x^A \tag{50}$$

As a final result consider the determinant of eq. (28).

$$\det(_aD_x^A) = \det(P) \left(\prod_{i=1}^{n} \,_aD_x^{\lambda_i}\right) \det(P^{-1}) = \prod_{i=1}^{n} \,_aD_x^{\lambda_i} \tag{51}$$

The right hand side of eq. (51) is a sequential fractional derivative (see e.g. [8] and [9]). Thus a sequential factional derivative may be viewed as the determinant of a matrix order fractional derivative. Additionally if the eigenvalues of A are such that $Re(\lambda_i) \leq 0$ then equation (51) can be simplified to give,

$$\det(_aD_x^A) = {_aD_x^{Tr(A)}} \qquad (52)$$

where Tr(A) is the trace of the matrix A, i.e. the sum of the eigenvalues.

## 6. Conclusion

In this brief report the Reimann-Liouville definition of fractional derivatives is used to obtain formulas for matrix order integration and differentiation. The resulting formulae were found to be well defined for any square matrix over the complex numbers. It is hoped that this new formalism will be of use in the study of systems of fractional differential equations and sequential fractional differential equations.

**Acknowledgment:** The author would like to thank P. Dorcey for helpful comments and a critical reading of the paper.